# The KNN rollercoaster: from bulk ceramics to phase engineered wafer-scale thin films

Giulia Pavese[1], Federico Orlando[1], Fabio Melzi[2], Walter Piazzi[1], Andrea Pescarolo[1], Federico Maspero[1], Marco Asa[1], Riccardo Gianola[2], Andrea Picco[2], Andrea Serafini[2], Kui Yao[3], Silvia Picozzi[4], Laura Castoldi[2], Miguel-Ángel Badillo-Ávila[1,*], Riccardo Bertacco[1,*]

[1]Department of Physics, Politecnico di Milano-Polifab, Via Giuseppe Colombo 81, 20133, Milan, Italy.

[2]STMicroelectronics, Via Camillo Olivetti 2, 20864, Agrate Brianza, Italy.

[3]Institute of Materials Research and Engineering (IMRE), Agency for Science, Technology and Research A*STAR, Singapore.

[4]Department of Materials Science, University of Milano-Bicocca, Via Roberto Cozzi 55, 20126, Milan, Italy.

[*] m.badillo.avila@outlook.com, [*] riccardo.bertacco@polimi.it

Keywords: lead-free piezoelectrics, ferroelectrics, $K_xNa_{1-x}NbO_3$ thin films, co-sputtering, wafer-scale deposition

**Abstract**

Since the initial disclosure of the extraordinary piezoelectric coefficients of Potassium sodium niobate (KNN) in near-equimolar bulk ceramics, its development trajectory has resembled a rollercoaster, with its integration into microelectronics severely lagging due to thermodynamic stability issues and poor planar process compatibility. In this work, we revisit the bulk-derived phase diagram for the specific case of thin films integrated on silicon. By systematically investigating Mn-doped $K_{1-x}Na_xNbO_3$ films grown on 8-inch wafers, we demonstrate that the optimal stoichiometry for thin films fundamentally diverges from the bulk equimolar standard. A Na-rich composition (> 70 at.%) is required to overcome substrate-induced constraints, effectively suppressing pyrochlore formation and chemical phase segregation while promoting dense columnar growth with a complete {001} out-of-plane polar orientation. Consequently, Na-rich films deliver outstanding functional properties, reaching remanent polarizations up to ~14 μC cm$^{-2}$, with piezoelectric coefficients of $d_{33,f} \approx 79$ pm V$^{-1}$ and $e_{31,f} \approx 10$ C m$^{-2}$. Supported by Density Functional Theory simulations, we correlate this enhancement with improved stability and a strain-driven structural reorientation toward a lower-symmetry monoclinic phase with tilted polarization. By redefining the phase engineering rules for wafer-scale thin films, our results establish a clear route toward KNN integration in microsystems.

## 1. Introduction

Since 2003, the European Union has sought to reduce the use of hazardous substances such as lead, mercury, and cadmium in electronic devices through the Restriction of Hazardous Substances (RoHS) Directive. Driven by the December 2026 expiration of the EU RoHS exemption for lead-based piezoelectrics like $Pb(Zr,Ti)O_3$ (PZT), the acceleration of the development of lead-free alternatives has become a critical regulatory and environmental priority.[1] For years, $K_{1-x}Na_xNbO_3$ (KNN) has been regarded as a promising lead-free alternative to piezoelectric lead-based materials used in a myriad of electronic devices.[2–5] Early work focused intensively on KNN bulk ceramics produced by conventional sintering. The landmark contribution by Saito et al. demonstrated very high piezoelectric coefficients in $K_{0.5}Na_{0.5}NbO_3$ ceramics and in films grown on engineered templates, stimulating an intensive research on this material.[6–8] Despite these advances, industrial adoption has been limited by stability and process-compatibility issues associated with volatile alkali elements (K, Na).[9–11] Nevertheless, KNN remains an attractive candidate for applications requiring high piezoelectric coefficients and full biocompatibility. Nowadays, one company offers 8-inch KNN-coated wafers for actuator and sensor applications and has announced a pilot line for piezo-MEMS fabrication.[12,13]

In piezoelectric oxides like PZT, phase boundaries (PB) between regions of different crystallographic symmetries are considered of particular interest. The coexistence of multiple phases typically leads to a softened crystal lattice, enhanced polarization rotation, and domain wall motion, resulting in improved piezoelectric properties.[14,15] Phase boundaries can be morphotropic (MPB), when the transition is induced by a small change in the composition or pressure conditions, or polymorphic (PPB), when the transition is strongly induced by temperature.[15,16] The most known and characterized MPB for perovskite PZT piezoelectrics occurs at $x = 0.48$ in $Pb(Zr_{(1-x)}Ti_x)O_3$.[17–19]

KNN is a solid solution between $KNbO_3$ and $NaNbO_3$ and can assume different phases, depending on temperature, composition[22], and doping.[14,23] Dissimilar to the phase diagram of PZT, that of KNN is more complex, as KNN can be found, in orthorhombic, tetragonal, rhombohedral, monoclinic and cubic phases with varying symmetries.[22] The original phase model for bulk KNN proposed by Ahthee and Glazer identified several phases, distinguished by their oxygen-octahedral tilt systems (Figure 1) and cation-displacement patterns.[24] The model was updated in the following years with several other studies at different temperatures and KNN compositions.[22,25,26] For $K_{1-x}Na_xNbO_3$, various works suggest the presence of a morphotropic phase boundary at $x$ around 0.5-0.55, between the K-rich orthorhombic phase and the Na-rich monoclinic phase.[27–29] Bulk KNN ceramics have indeed demonstrated maximum piezoelectric coefficients and dielectric constants at $x$ around 0.5.[28,30–32] As suggested by Levin et al. [27], such enhancement is driven by pre-transitional, short-range octahedral rotations that soften local bonds and reduce the energy barrier for polarization rotation. Although the morphotropic phase boundary at $K_{0.5}Na_{0.5}NbO_3$ has been the most extensively studied in KNN ceramics, other two phase boundaries suggested by phase diagrams in the Na-rich region may also provide enhanced piezoelectric performance.

According to Baker et al.,[29] KNN further undergoes phase transitions at high Na content, around $x = 0.65$ and $x = 0.8$, mainly associated with changes in the octahedral tilt system. Using Glazer notation,[21] (see Supporting Information, Section S1) the non-tilted monoclinic phase ($a^0b^0c^0$) is stable between $x = 0.5$ and $x = 0.65$, while at $x \simeq 0.65$ a transition to the $a^0b^+c^0$ configuration occurs, characterized by in-phase tilts around the $b$ pseudo-cubic axis. At higher Na content (x $\simeq$ 0.8), the structure evolves toward the more complex tilt pattern $a^-b^+c^-$, involving additional out-of-phase rotations of neighboring octahedra around the $a$ and $c$ axes. Indeed, in bulk KNN, compositions near the second (x ~ 0.7) and third (x ~ 0.83) phase boundaries have shown promising properties, including high remanent polarization and piezoelectric coefficients in both undoped ceramics and Mn-doped single crystals.[33–35]

Motivated by the improved properties reported for bulk KNN ceramics near the equimolar Na/K composition ($K_{0.5}Na_{0.5}NbO_3$), most studies on thin KNN layers have likewise focused around this particular compositional region. Therefore, only a limited number of authors have explored a wider compositional range. Among such, Shibata et al. identified the highest $d_{31}$ at Na/(Na+K) = 0.55 in sputtered KNN films of various compositions (the corresponding $d_{33}$ and $P_r$ values were not disclosed).[10] In a later study by the same authors, $K_{0.34}Na_{0.66}NbO_3$ films (close to the second phase boundary) exhibited even better performance; however, the rationale for such compositional choice was not clarified.[12] Likewise, Kanno et al.[37] reported sputtered KNN films near the third phase boundary with high $P_r = 25$ µC cm$^{-2}$ and relatively low $e_{31,f} = 2.7$ C m$^{-2}$. Interestingly, in a subsequent conference paper by the same group, it was further showed that Na-rich films exhibit a more columnar morphology and stronger (001) diffraction intensity than K-rich films, pointing to improved film quality near the Na-rich phase boundaries.[38]

In this rollercoaster of KNN adoption, which is now moving from laboratory studies toward wafer-scale films and production, important fundamental questions remain unresolved, particularly for thin films, for which substrate-induced strain may substantially modify the phase diagram.[36] The present work addresses one critical issue: identifying the optimal stoichiometry that balances chemical/thermal stability with maximal piezoelectric performance in KNN films. We investigated the co-sputtering growth of $K_{1-x}Na_xNbO_3$ thin films with increasing Na/K > 1 ratios on 8-inch wafers. We demonstrate that high-performing KNN films with optimized ferroelectric and piezoelectric properties can be achieved at compositions close to the second and third phase boundaries, respectively. The various structural and electrical analyses, combined with Density Functional Theory (DFT) simulations, reveal that the superior piezoelectric properties observed at high Na at.% may be associated with phase transitions toward lower-symmetry monoclinic structures that might facilitate polarization rotation, highlighting a complex interplay between substrate-induced strain and composition-driven phase transitions.

## 2. Results

$K_{1-x}Na_xNbO_3$ ($0.48 < x < 0.83$) films with different Na/K ratios were grown by RF, confocal co-sputtering on 8-inch Si-wafer substrates using an industrial EVATEC Clusterline®-200E tool. Depositions were performed using various 4-inch $K_{1-x}Na_xNbO_3$ (including $x = 1$ and $x = 0$) targets doped with a small amount of Mn. The targets were developed on-request. All films were grown in a mixed Ar + $O_2$ atmosphere at nominal 600 °C on Pt(111)/$TiO_2$/$SiO_2$/Si(100) templates. Compositional control of KNN films was achieved by varying the RF power applied to the various single RF sources during co-sputtering. Top electrodes on films consisted of a sputtered 100 nm-thick Pt layer. The KNN-stack wafers were cut into smaller 2x2 cm coupons and annealed in a Unitemp-RTP furnace at 650 °C for 30 min under an oxygen atmosphere.

| **Sample** | **Thickness [nm]** | **Estimated Na at.%** |
|---|---|---|
| KNN-42 | 725 | 41.6 ± 0.9 |
| KNN-48 | 854 | 47.7 ± 1.1 |
| KNN-61 | 830 | 60.8 ± 1.0 |
| KNN-66 | 785 | 66.0 ± 0.9 |
| KNN-68 | 960 | 68.0 ± 2.3 |
| KNN-73 | 755 | 73.5 ± 0.8 |
| KNN-78 | 835 | 78.1 ± 0.9 |
| KNN-83 | 790 | 82.8 ± 0.9 |

**Table 1**. EDX estimation of the composition of KNN films. Several elemental maps were recorded and analyzed to improve the reliability of the data.

Table 1 shows the percentage of Na/(K+Na) ratios for all the KNN compositions measured by EDX (Electron Dispersive X-ray) spectroscopy. For ease of discussion, the Na percentages have been included in the names of samples as digits after the acronym "KNN". Considering the importance of proximity to phase boundaries compositions for KNN, our analysis focuses mainly on three samples: KNN-48, KNN-68, and KNN-78.[24,29]

**2.1. Functional properties of co-sputtered KNN films**

Figure 2a shows the typical KNN capacitor stack used in the various methods of characterization. To assess the ferroelectric behavior of our KNN films, Dynamic Hysteresis Measurements (DHM) were performed by applying triangular voltage pulses at 5 kHz to the Pt top electrodes. The maximum voltages applied were adjusted to the thickness of the films to reach an $E$ = 250 kV cm$^{-1}$. The piezoelectric $d_{33,f}$ and $e_{31,f}$ coefficients were, instead, assessed by Double Beam Laser Interferometry (DBLI) using the technique discussed in the methods section. Dielectric constant ($\varepsilon_r$) values were estimated by C-V measurements.

Figure 2b shows the electrical current loops for KNN-48, KNN-68, and KNN-78 films, from which the polarization loops are derived. All three compositions display clear ferroelectric switching currents and polarization curves, with increasing coercive fields around 40 kV cm$^{-1}$ for KNN-48, 43 kV/cm for KNN-68 and 46 for KNN-78, which align well with values reported in the literature.[37,38] In contrast to our previous work, the present KNN films did not require leakage compensated measurements. For most KNN compositions, leakage current densities were reduced to less than 10$^{-6}$ J cm$^{-2}$ at 400 kV cm$^{-1}$. Along with improvements in deposition and processing conditions, a major electrical enhancement occurred after the addition of small quantities of Mn to the targets. The use of Mn as doping agent of KNN has been reported by several authors.[39–43]

According to various studies, Mn can occupy either the A-site or the B-site in the KNN perovskite lattice and can compensate p-type and n-type conduction.[40,41,44,45] The volatilization of alkali ions promotes the formation of alkali vacancies, which act as acceptor defects and contribute to hole generation, as well as oxygen vacancies, which behave as donor defects and contribute to electronic conduction, thereby preserving charge neutrality. Oxygen vacancies may be further oxidized during annealing. $Mn^{2+}$ can compensate holes by donating electrons and, consequently, increasing its oxidation state to $Mn^{3+}$.[43] According to the model proposed by Phu et al., $Mn^{2+}$ introduced through Mn–O doping may incorporate at the A-site, releasing an oxygen ion that helps compensate nearby oxygen vacancies. Alternatively, $Mn^{2+}$ may substitute for $Nb^{5+}$ at the B-site, increasing its valence to $Mn^{3+}$ or $Mn^{4+}$ and compensating hole carriers through the same mechanism described above.[45]

The dielectric loss tangent was measured at room temperature and 5 kHz, resulting in values of $\tan\delta \approx 0.04–0.06$ at zero bias field. Our KNN films display slightly higher losses than highly optimized ($\tan\delta \approx 0.017$) or epitaxial KNN thin films ($\tan\delta \sim 0.02-0.04$) reported in the literature[46,47], yet they remain fully comparable to values commonly reported for Pt/KNN/Pt capacitor structures exhibiting strong ferroelectric and piezoelectric responses.[13] Piezoelectric measurements carried out by DBLI are reported in Figure 2c, showing the characteristic butterfly strain loops when sweeping electric field, from which the values of the piezoelectric coefficients have been extracted.

Noteworthy, both ferroelectric and piezoelectric properties of our KNN films exhibit a dependence on the Na composition of each sample. The trends are displayed in Figure 2d and Figure 2e. Regarding ferroelectric behavior, a maximum mean value of $P_r$ = 13.2 $\mu$C cm$^{-2}$ is obtained for the KNN-68 sample (Figure 2d, red curve) near the 2$^{nd}$ phase boundary ($PB_2$), while the coercive field slightly increases with higher Na loads and reaches a maximum mean $E_c$ = 46.4 kV cm$^{-1}$ for KNN-78 (Figure 2d, purple curve). The values for the remanent polarization are in line with others already reported in the literature for Mn-doped KNN.[12,48] Concerning the piezoelectric response, KNN-78 achieves the highest coefficients of $d_{33,f}$ = 78.8 pm V$^{-1}$ (Figure 2d, blue curve) and $e_{31,f}$ = 10.2 C m$^{-2}$ (Figure 2e, green curve) coinciding with the vicinity of the 3$^{rd}$ phase boundary ($PB_3$). The high values of the piezoelectric coefficients are also in line with those of KNN films reported in the literature.[12,49] Notice that, the $d_{33,f}$ and $e_{31,f}$ coefficients follow a similar improvement trend, as $P_r$, with respect to Na at.%. However, maximum polarization and mechanical responses are obtained for slightly different Na compositions of 68% and 78%, respectively. This is not totally unexpected, as the ferroelectric polarization and the piezoelectric coefficients are two different physical entities, but it constitutes a first indication of the fact that the polarization vector at high Na concentration might not be aligned with the out-of-plane direction of application of the electric field.

Dielectric constant values ($\varepsilon_r$) were calculated from capacitance-voltage measurements at zero field; the values are presented in Figure 2f. The data points demonstrate a net increase of $\varepsilon_r$ going from KNN-42 to KNN-83. The low value of $\varepsilon_r$ for KNN-42 can be easily ascribed to a mixture of low-quality perovskite and presence of spurious pyrochlore phases. Dielectric constants of 800, and above, are obtained for higher Na compositions, suggesting a much higher quality of the perovskite material. Interestingly, local maximums in $\varepsilon_r$ are found for KNN-48,

KNN-68, and KNN-78, which are the compositions close to the reported phase boundaries for KNN ceramics (see Figure 1).

In PZT-based materials, a maximum of the dielectric constant, together with a maximum of the piezoelectric properties, has been observed at the MPB between different phases. The reason of the abrupt increase has been attributed to the presence of several polarization states, originating from the co-existence of tetragonal and rhombohedral phases at MPB, which have nearly equivalent free energy. This provides a high polarization for the MPB composition, which in turn favors strong dielectric and piezoelectric effects.[50–52].

While for KNN solid solutions a peak in the dielectric constant has been reported for the first phase boundary, the trend has not been universal. Kumar et al. [53] found that $K_{0.5}Na_{0.5}NbO_3$, where orthorhombic and monoclinic phases coexist, showed a higher dielectric constant than nearby compositions, suggesting that phase coexistence may enhance dielectric properties. In contrast, Dai et al.[54] identified the phase boundary at 52 at.% Na and observed a maximum in $d_{33}$, but not in the dielectric constant. Near the second MPB, Zhang et al.[30] observed a discontinuity in the lattice parameter and maxima in $d_{33}$. Meanwhile, the dielectric constant reached its peak between 40 and 60 at.% Na. In general, comprehensive data regarding the third phase boundary remains scarce in the current literature.

To summarize, enhanced functional properties were observed for KNN-68 and KNN-78 films, where both the remanent polarization and the piezoelectric coefficients reached their maximum values. The dielectric constant showed a local maximum for KNN-48 and KNN-78 and a global maximum for KNN-68. Notably, these compositions are located near the reported phase boundaries (PB) for bulk KNN ceramics. In the following section we focus on the structural and chemical characterization of our film samples, to elucidate the nature of these phase boundaries in oriented films and clarify the reason why a specific phase boundary should be preferred for application of KNN in integrated devices.

**2.2. Dependence of morphology and local composition on alkali stoichiometry**

Figure 3a, b, and c, show SEM cross sections of films for three KNN compositions close to the expected bulk phase transitions (48, 68, and 78 at.% Na). All the films exhibit a packed columnar structure, which is commonly reported for KNN and other high quality piezoelectric films with preferential out of plane orientation.[55,56] Sample KNN-48 (Figure 3a and d) exhibits a fine columnar assembly; its surface morphology is relatively flat with clearly defined grain boundaries, though it contains some defects in the form of crevices and small particles. The KNN-68 film (Figure 3b and e) maintains a columnar texture; however, both the columns and grains look noticeably larger. Average grain size diameters were calculated with Gwyddion software.[57] An increase from 157 ± 40 nm for KNN-48, to 224 ± 90 nm for KNN-68 was estimated from top-surface SEM images. As the sodium atomic percentage increases further, the columnar structure of KNN-78 films improves (Figure 3c), resulting in a less evident grain separation (Figure 3f), and a similar grain size (171 ± 46 nm). Additionally, protruding crystallites (emphasized by green arrows in Figure 3c and f) appear. In other reports, such geometric crystallites have been linked to individual KNN grains with (111) or (110) orientations.[49]

Figures 3g, h, and i, show elemental maps of Na, K and Nb for KNN-48, KNN-68, and KNN-78 films obtained from energy-dispersive x-ray spectroscopy (EDX) measurements of the various lamellas. The compositional maps evidence some differences for the three KNN compositions analyzed. While Nb seems well distributed, for K and Na the situation is different. In all films, there are broad areas in which alkali ions are more evenly distributed (exemplified by orange circles in the figures). Nonetheless, some highly localized segregation is also found in the various films (white circles in the images). For KNN-48 (Figure 3g), Na tends to segregate locally into some concentrated regions, while at the same time K is less present in such spots. A similar situation happens for KNN-68 (Figure 3e). The case of KNN-78 is distinctive. Regions with higher Na/K ratio appear sharply defined, they increase in size and occur more frequently across the film. At the surface, these regions adopt the morphology of triangular crystallites protruding from the film, as highlighted by the green arrow in Figure 3f. The elemental maps were analyzed locally, and quantitative results are presented in Table 2.

| **Sample** | **Region** | **Na/(Na+K)** | **(Na+K)/Nb** | **O/Nb** |
|---|---|---|---|---|
| KNN-48 | Na-rich | 0.80 ± 0.03 | 0.72 ± 0.09 | 2.73 ± 0.09 |
| KNN-48 | Even | 0.49 ± 0.04 | 0.78 ± 0.09 | 2.66 ± 0.14 |
| KNN-68 | Na-rich | 0.82 ± 0.02 | 0.93 ± 0.04 | 3.00 ± 0.08 |
| KNN-68 | Even | 0.70 ± 0.07 | 1.00 ± 0.05 | 3.07 ± 0.11 |
| KNN-78 | Na-rich | 0.89 ± 0.02 | 0.95 ± 0.09 | 3.02 ± 0.08 |
| KNN-78 | Even | 0.78 ± 0.02 | 0.95 ± 0.09 | 2.96 ± 0.20 |

**Table 2**. EDX compositional estimations of localized Na-rich and even areas in horizontal lamellas from the middle of the films.

The evenly distributed regions display a Na/K composition closely corresponding to the average one evaluated by EDX on the top surface, as reported in Table 1. KNN-48 shows some clear alkali deficiency, while samples with higher Na content show an almost ideal stoichiometry with (Na+K)/Nb ratios close to 1. This is not surprising due to the higher volatility of potassium and the intrinsic poor stability of the 50/50 stoichiometry.[11] The sodium rich spots show instead a Na/(K+Na) ratio slightly increasing from ~ 0.80 to ~ 0.89 when moving from KNN-48 to KNN-78. At variance with our previous study on Na segregation in $K_{0.5}Na_{0.5}NbO_3$ films, however, the higher concentration of sodium in some spots is not accompanied by a local excess of alkali. As a matter of fact, the (K+Na)/Nb ratio remains close to 1, and the O/Nb ratio is very close to 3, thus confirming that a good perovskite develops also in these areas with typical size between 0.1 and 1 μm. This is a clear confirmation of the tendency of the system to promote the formation of stable Na-rich phases with Na composition around 80 at.%, but in our optimized growth conditions this phenomenon occurs at the sub-micron scale. Such behavior is highly different from the large Na-rich islands observed in our previous paper (up to 70 μm).[11]

Overall, films with nominal Na content above 70 at.% display a superior chemical stability and the full formation of a perovskite phase with the proper ratio between alkali, Nb, and oxygen. This is a fundamental point supporting the fact that optimum piezoelectric and ferroelectric properties can be obtained not at the usual 50/50 Na/K ratio reported for ceramics, but in Na-rich solid solutions in the case of thin films.

**2.3. Structural and orientation analysis**

Figure 4a and Figure 4b show x-ray diffractograms of the KNN films for three Na/(K + Na) ratios acquired in out-of-plane and in-plane measuring geometries, respectively. Because of extensive phase segregation, KNN-42 lattice parameters were omitted from the trends. In Figure 4a, the out of plane measurements exhibit dominant reflections at $2\theta \approx 22.5°$ and 45.5°, corresponding to the {001} family, thus evidencing a strong (001) orientation out of plane of the pseudo-cubic KNN perovskite structure.[37] The use of a logarithmic scale makes it evident that other orientation reflections are virtually absent in the out of plane direction. The strong pseudo-cubic (001) preferred orientation, perpendicular to the substrate, was confirmed through polar measurements for KNN-68. Figure 4c shows fused polar plots for KNN (001) and KNN (011) reflections. An intense dot at the center of the polar map ($\alpha$ = 0°) demonstrates that crystallites with (001) orientation are highly aligned to the vertical axis, with a FWHM of 3.4°. Instead scanning of the (011) reflection results in a texture ring around the vertical at a tilt angle $\alpha$ = 45º, indicating a semi-random in-plane azimuth with a FWHM of 4.5°. This "fiber texture" of crystallites is consistent with the columnar growth observed in the SEM micrographs and is highly desirable for piezoelectric applications.[37,59] The [001] direction represents the polar axis for the tetragonal symmetry of perovskite KNN (stable at high temperature). For the case of the orthorhombic symmetry (stable at room temperature), the polar axis is technically the $[110]_O$ direction because it is constructed with a tilting of the perovskite cell. Yet, $[110]_O$ is parallel to the [001] pseudo-cubic axis used as the common framework for comparison (indexed as $[001]_{pc}$). For this reason, the $[001]_{pc}$ vertical-growth is typically targeted during deposition of KNN films.[31]

Upon detailed analysis, data in Figure 4a shows that for KNN-48 a minor contribution of $(100)_{pc}$-oriented crystals is visible as a shoulder to the right of the main $(001)_{pc}$ peak. Nevertheless, as the sodium content increases beyond 60 at.%, this dual orientation is suppressed, yielding a high (001) texture. Additionally, as sodium concentration increases, a clear shift of the {001} reflections toward higher 2θ angles is observed, representing a contraction of the unit cell due to the smaller ionic radius of $Na^+$ compared to $K^+$.[60] The appearance of a KNN (110) reflection at 2θ ≈ 32.2° for KNN-73, and beyond 73 Na at.%, correlates well with some misoriented triangular facets observed via SEM (presented in the previous section).[49] The possible presence of a (111) out-of-plane orientation is, however, masked by the high intensity of the Pt (111) peak (~39.9°). Other than the perovskite phase, from KNN-42 to KNN-68 small reflections at ~ 29.3° and ~ 29.8° indicate the presence of spurious pyrochlore phases that typically form during the annealing process at around 600 °C.[11] In contrast, these reflections are virtually absent for KNN-72, KNN-78, and KNN-83. Thus, the experimental results demonstrate that higher Na-content is also instrumental in suppressing non-perovskite secondary phases.

Regarding in-plane diffractograms of the KNN films (Figure 4b), several reflections of perovskite KNN are present in all films. These include contributions from the {001}, {011}, {012}, and {111} planes. The relative intensities of such reflections do not correspond to a totally randomly oriented powder. The (111) plane, for example, has a particularly low relative intensity with respect to the case reported for KNN ceramics. This can be justified by the aforementioned "fiber texture", where the material orients preferentially in out of plane direction but semi-randomly in the in-plane direction. Although the in-plane measurements provide us with more structural information, peak deconvolution remains challenging due to a limited resolution of our standard Cu-based XRD equipment, and because of the close similarities between lattice parameters of the various KNN polymorphs.

The XRD data was used to estimate the lattice parameters of our KNN targets and films. In standard indexing for orthorhombic KNN in a pseudo-cubic framework, *b* is typically assigned to the shortest axis and the longer *c* axis is almost equal to *a* ($a > b$).[26,29] The long *c* and *a* lattice parameters are difficult to resolve for thin films with Cu-Kα radiation as their interplanar spacing is on the order of 0.02 Å, corresponding to an angular separation of only about $2\theta$ = 0.2°-0.25°. Hence, in the following discussion we will focus mostly on the long $c_{pc}$ and short $b_{pc}$ parameters (*pc* stands for pseudo-cubic), used as comparison for the longer and shorter axis found in KNN.

For the ceramic targets, Bragg-Brentano geometry was used for XRD measurements; $c_{pc}$ and $b_{pc}$ were calculated from $\theta$-$2\theta$ diffractograms considering the {001} and {100} families of planes. Peaks were fitted considering a pseudo-cubic perovskite model in Rigaku SmartLab Studio. The estimated lattice parameters decrease monotonically with increasing Na content. As mentioned earlier, the trend is consistent with the smaller ionic radius of $Na^+$, which leads to a progressive contraction of the unit cell. Moreover, the persistent difference between the pseudo-cubic parameters indicates that the structure is in agreement with an orthorhombic/monoclinic symmetry expected from the KNN phase diagram for powders (Figure 1). In general, the lattice parameters for the ceramic targets demonstrate a good agreement with the values reported in the literature for KNN ceramics at similar compositions.[22,26,58]

In contrast, thin films exhibited some behavior that was not observed for the targets and for bulk KNN ceramics. The data in Figure 4a and b was used to estimate the in-plane and out of plane lattice parameters for KNN films as a function of Na at.%. Results are presented in Figure 4d together with lattice parameters reported in the literature for ceramic KNN close to the three phase boundaries.[22,26,58] In this study, however, rather than long $c_{pc}$ and short $b_{pc}$, the lattice parameters for KNN films are denoted as $a_{OOP}$ (out-of-plane) and $a_{IP}$ (in-plane). Such nomenclature is adopted to avoid ambiguity, as the crystallographic $c$-axis and $a$-axis are not consistently aligned with the out-of-plane and in plane direction across all samples. In a similar fashion to the targets, the evolution of the lattice parameters shows a general reduction of both out-of-plane and in-plane distances with increasing Na content for the KNN films. Nevertheless, a clear distinction between $c_{pc}$ ($a_{OOP}$) and $b_{pc}$ ($a_{IP}$) was, in several cases, not possible.

As shown in Figure 4d, for compositions between 48 and 73 Na at.%, the $a_{IP}$ is generally the shorter axis; however, it is strongly elongated with respect to the values reported for the $b_{pc}$ in the literature, demonstrating a pseudo-cubic cell with $a_{OOP}/a_{IP}$ close to 1. This trend is particularly cumbersome above 48 at.% Na, where the two lattice parameters become almost indistinguishable from each other. Such apparent elongation of the in-plane lattice parameter can have two concurrent origins. On the one hand, the in-plane reflection may contain unresolved contributions from both the $a$ and $b$ lattice parameters; because the expected peak splitting can be smaller than the instrumental resolution and comparable to the peak broadening, the two reflections overlap and are detected as a single peak. On the other hand, it might be an effect of a tensile strain state in the film, imposed by thermal and lattice mismatch with the substrate. Unlike bulk materials, which are typically free from external mechanical constraints, thin films are laterally clamped by the substrate, leading to significant lattice distortions. An initial estimation of macro-strain in the KNN films was obtained from profilometer measurements using Stoney's method, which allows for the strain to be evaluated by measuring the radius of curvature of wafers caused by thin film stacks deposited on them.[61] The result evidenced that, after growth and annealing, our KNN films are subjected to tensile strain with magnitudes of around +450 MPa. This observation suggests that tensile strain imposed by bottom layers in the stack contributes, at least in part, to the enlarged in-plane lattice parameter observed in our KNN films.

More interestingly, a critical transition occurs above 73 at.% Na, where a significant decrease in $a_{OOP}$ is observed leading to a $a_{OOP}/a_{IP} < 1$, possibly indicating a phase transition or either a 90° structural reorientation. Beyond this composition, $a_{OOP}$ approaches the $b_{pc}$ value reported for KNN ceramics, while $a_{IP}$ becomes closer to the corresponding $c_{pc}$ value. Thus, above the 73 Na at.% threshold, the structure deviates from the more strained pseudo-cubic symmetry (where lattice parameters are closer to each other) and instead approaches a monoclinic or orthorhombic phase with a more pronounced separation between the lattice parameters. In this regime, the longer lattice parameter is likely forced in-plane by the strong tensile stress imposed by the substrate. Improvement in piezoelectric properties of KNN has been reported in the past for films satisfying $a_{OOP}/a_{IP} < 1$.[62]

## 2.4. DFT studies

To complement the experimental scenario, we also performed a systematic Density Functional Theory (DFT) study of the monoclinic $Pm - a^0b^+c^0$ phase of KNN across different compositions. This particular phase was selected as the target structure because it occupies a central position in the Na-rich side of the bulk phase diagram (Figure 1), thus better matching most of the KNN compositions investigated experimentally. At the same time, it represents a convenient compromise in terms of computational overload. On one hand, it exhibits a non-trivial, yet not excessively intricate, structural complexity, being intermediate between the non-tilted $Pm - a^0b^0c^0$ phase and the fully tilted $Pm - a^-b^+c^-$ phase. On the other hand, it allows for a relatively small primitive unit cell containing only 10 atoms, which is crucial for constructing simulation cells of controlled size. It is important to note that all DFT simulations were performed considering free-standing cells, i.e., without any external strain imposed by a substrate framework (such as Pt(111), employed in the experimental samples). Therefore, the following analysis is meant to provide an initial framework to investigate the effect of Na composition in KNN films, without including the additional constraints associated with growth on a substrate.

The chemical disorder on the A-site sublattice of KNN was modeled by means of the Special Quasirandom Structure (SQS) approach, implemented in a suitably chosen supercell (further details are provided in the Experimental section).[11] We sampled a wide compositional range, overlapping with the experimental data points as closely as possible, namely, considering the concentrations $x$ = 0.50, 0.67, 0.71, 0.75, 0.79, and 0.83. For completeness, we also included the end-member compositions $x = 0$ (pure $KNbO_3$) and $x = 1$ (pure $NaNbO_3$) as reference points, although these were not investigated experimentally. Figure 5a illustrates a representative subset of the relaxed SQS supercells in the *xz* polar plane. Octahedral tilting increases across the compositional series, although not strictly linearly and in a locally dependent manner, since tilt amplitudes are governed by the quasirandom distribution of Na and K atoms on the A sites surrounding each individual octahedron.

To start with, we evaluated the mixing enthalpy ($H_{mix}$) to assess stability with respect to phase separation across the composition range. $H_{mix}$ compares a homogeneous alloy at a given composition ($x$) to a macroscopic mixture of pure $KNbO_3$ and pure $NaNbO_3$, by measuring the deviation of the alloy energy $E(x)$ from a linear interpolation between the energies of the two end members: $H_{mix}(x) = E(x) - [(1 - x)E(0) + x\,E(1)]$. As shown in Figure 5b, $H_{mix}$ is positive throughout the entire compositional range, reaching values of approximately 40 meV, about an order of magnitude larger than those obtained enforcing tetragonal symmetry (≈ 5 meV), as reported in our previous work.[11] This confirms (and strengthens) the tendency of KNN to be unstable with respect to phase separation into the pure end members, particularly at compositions around $x = 0.7$. This is also coherent with the observation of sub-micrometric segregation in KNN films for all compositions, reported in Section 2.2. However, in the experimental scenario, the formation of Na-rich phases seems to be favored with respect to K-rich phases.

We then explored the compositional dependence of the $d_{33}$ coefficient for the $Pm$–$a^0b^+c^0$ phase. As clarified in Figure 5a, the lattice was oriented with cell vectors $A_1$ and $A_3$ at 45° to the global $x$ and $z$ axes to match the experimental configuration observed for Na-rich films, where the out-of-plane measurement direction ($z$) does not align with the polar axis. Because DFT piezoelectric tensors are computed in the global frame, the reported $d_{33}$ coefficient represents the piezoelectric response along the global *z*-axis (experimental). Thus, the calculated $d_{33}$ provides a direct theoretical counterpart to the experimental out-of-plane measurements, as both capture the response along a non-polar axis

Figure 5c reports the calculated $d_{33}$ values, compared with the experimental data, in the range $x \simeq 0.66$–0.83, where the $Pm - a^0b^+c^0$ phase is expected as per the bulk phase diagram. Agreement is observed only within the compositional window $x \simeq 0.75$–0.83, where the absolute values nearly coincide and the local maximum predicted by DFT reproduces the global maximum observed experimentally. The discrepancy between experimental results and DFT predictions at lower Na compositions is not surprising, as we do not find clear evidence of monoclinic symmetry in films with $x < 0.7$, since they display a pseudo-cubic symmetry (see Section 2.3). On the other hand, Figure 5c suggests that the monoclinic phase investigated here may indeed be present in the experimental samples for $x > 0.7$, consistent with the indications provided by XRD data. Although the agreement is limited to a relatively narrow compositional range, it is worth emphasizing that this interval lies approximately within the proposed stability region of the monoclinic $Pm - a^0b^+c^0$ symmetry, which in the present experimental scenario is likely influenced by substrate-induced strain. Furthermore, the (a priori arbitrary) structural setting adopted in our calculations – with a polar axis rotated by approximately 45° away from the *z* direction – may in fact reasonably approximate the effective configuration realized in the experimental samples. As discussed in the previous sections, both ferroelectric and XRD measurements indicate that for $x > 0.7$ the polarization vector (i.e., the polar axis) is not aligned with the direction of maximum elongation of the crystal unit cell.

## 3. Discussion

Eight Mn-doped $K_xNa_{1-x}NbO_3$ (KNN) films with Na contents between 42 and 83 at.% were grown by co-sputtering on 8-inch wafers to systematically explore the influence of Na content on the properties of the films. Cross-sectional SEM images revealed a columnar morphology typical of (001) fiber-textured films for all KNN compositions, while top-view imaging shows a relatively flat surface for compositions between 48 and 73 Na at.%. KNN films with high Na concentrations (> 78 at.% Na) start to develop crystallites protruding from the film surface. STEM–EDX mapping revealed sub-micrometer alkali segregation in all compositions, leading to the formation of Na-rich regions that, nonetheless, still display nominal $ABO_3$ perovskite stoichiometry. The observations indicate that, even in optimized growth conditions, the KNN system retains a natural tendency to segregate into distinct phases. However, macroscopic segregation and structural incoherence can be reduced at sufficiently high Na content, resulting in functional films with suitable functional properties appropriate for device applications. These results are consistent with our previous work, where the higher volatility of potassium was shown to promote Na enrichment and the formation of Na-rich perovskite phases. In KNN films grown by large-scale confocal sputtering, this led to the formation of Na-rich islands ($K_{0.3}Na_{0.7}NbO_3$–$K_{0.2}Na_{0.8}NbO_3$), which were identified as the origin of the macroscopic ferroelectric response.[11] Overall, this supports the enhanced stability of Na-rich phases in the KNN film system.

The electrical, ferroelectric and piezoelectric results evidenced a strong dependence on stoichiometry: the highest remanent polarization was observed at 68 at.% Na, close to the second phase boundary of bulk ceramic KNN samples, while the piezoelectric response peaked at 78 at.% Na, close to the third phase boundary, reaching $d_{33,f}$ = 78.8 pm V$^{-1}$ and $e_{31,f}$ = 10.2 C m$^{-2}$. This can be explained in terms of a structural transition from a weakly distorted pseudo-cubic phase for $x < 0.7$, to a real low-symmetry monoclinic phase for $x > 0.7$. The structural change implies reorientation of the polarization vector from the out of plane direction (for $x < 0.7$) to an oblique direction typical of a monoclinic or orthorhombic phase. By consequence, and due to the projection of the polarization vector along the surface normal, we expect to experimentally measure a smaller value of the polarization in KNN micro-capacitors.

X-ray diffraction analysis supports this scenario. Overall, lattice parameters contract with increasing Na, consistent with its smaller ionic radius. As a side effect, increasing Na content progressively suppresses pyrochlore formation. More importantly, a preferential $(001)_{pc}$ out-of-plane orientation was initially observed for films with Na content < 73 at.%. The highest tetragonality in the out-of-plane direction being detected for the KNN film with ~48 Na at.% (close to the first phase boundary of KNN); whereas a more cubic-like structure was observed at higher Na contents. Above ~73 Na at.%, however, an inversion of the out-of-plane and in-plane experimental lattice parameters ratio ($a_{OOP}/a_{IP} < 1$) was clearly found, indicating a strain-driven structural reorientation, and thus a preferential $(100)_{pc}$ orientation out of plane. Despite the in-plane orientation of the longer axis, a non-zero out-of-plane remanent polarization was measured in KNN-78 and KNN-83. For a strictly tetragonal symmetry with the long $c$-axis in plane (and to an extent for the orthorhombic system also), tensile strain would confine the polarization vector entirely (or mostly) to the film plane, yielding near zero out-of-plane polarization. The observation of measurable polarization in in-plane $c_{pc}$-axis oriented films indicates that the unit cell displays a lower-symmetry orthorhombic and/or monoclinic configuration, allowing the polarization vector to deviate from the primary out of plane axes.

Considering this aspect, and the expected monoclinic phase that is observed in the phase diagram for bulk KNN[22,58] for corresponding Na concentrations, we can hypothesize that our Na-rich KNN films (KNN-78 and KNN-83) might possess monoclinic symmetry or be in mixtures with the orthorhombic. The proposed structural arrangement would facilitate the polarization rotation mechanism, as a tilted polarization vectors respond piezoelectrically more strongly to external electric fields than rigidly aligned ones.[63] The model is supported by the maxima observed in the $d_{33,f}$ and $e_{31,f}$ piezoelectric coefficients for KNN-78, reaching values up to 78 pm $V^{-1}$.

The enhancement might be similar to the behavior observed at the morphotropic phase boundary of PZT, for which intermediate monoclinic states between tetragonal and rhombohedral phases have been observed.[64] DFT calculations for KNN further support this interpretation, as a good matching between experimental values of the $d_{33}$ coefficient and the theoretical ones, evaluated by fixing a monoclinic structure and varying the Na concentration, are found only for $x > 0.7$, where we observe an evident structural transition from XRD data. Another small, but relevant observation, is that the dielectric constant of the films exhibited local maxima at 48, 68, and 78 at.% Na. This vicinity of enhanced functional properties to compositions identified as phase boundaries in the literature might suggest that, in our KNN films, there might be hints of structural transitions at those compositions, like the ones observed in the literature for bulk KNN systems. Unfortunately, the experimental techniques employed in this work did not allow for unambiguous individual determination of crystal structures. Further clarification could be achieved through synchrotron-based diffraction measurements or high-resolution scanning transmission electron microscopy (HRTEM); the latter is currently being pursued in ongoing work to provide deeper insight into this aspect. Eventual further results will be discussed in a future publication.

Beyond a possible phase change, films with higher Na atomic percentage exhibit improved alkali-ion retention and a more effective suppression of the pyrochlore phase.

A final consideration should be made about the optimum stoichiometry to be chosen for KNN films. Our results underscore the role of the substrate in determining the effective phase diagram of the films, mainly through strain. This means that substrate and strain engineering should be carefully considered with reference to the specific substrate/application.[62,65,66] On the other hand, the general thermodynamic tendency to film segregation in order to stabilize Na rich regions definitely suggests avoiding the "optimum" $K_{0.5}Na_{0.5}NbO_3$ stoichiometry for bulk ceramics and move to Na-rich solid solutions close to the second and third phase boundaries of the KNN bulk phase diagram.

## Conclusions

In the KNN rollercoaster, our work can help in guiding the steps of material development towards the effective integration of KNN in the next generation of lead-free piezoelectric devices. This work demonstrates that the best ferroelectric properties in our KNN films are obtained at 68 at.% Na, whereas the highest piezoelectric coefficients are achieved at 78 at.% Na, indicating that the optimal stoichiometry in thin films is shifted toward Na-rich compositions compared to the equimolar composition typically considered optimal for bulk ceramics. Moreover, Na-rich KNN displays higher stability against loss of alkali elements, better control of {001} orientation, improved cross-section columnar morphology, and reduced presence of pyrochlore and segregated phases. This behavior is associated with energetical considerations addressed by DFT, and a structural evolution from a weakly distorted pseudo-cubic phase, at lower Na contents, to a lower-symmetry phase at higher Na compositions accompanied by a reorientation of the lattice parameters. For Na-rich KNN films, such structural configuration likely favors polarization rotation and results in enhanced piezoelectric properties. Our results suggest that substrate-induced strain plays a key role in defining the effective phase stability and orientation of KNN thin films and point to Na-rich compositions near the second and third bulk phase boundaries as the most promising for piezoelectric microdevice applications.

## Experimental Section/Methods

For crystallographic characterization, we utilized a Rigaku SmartLab XE X-ray diffraction system, equipped with a copper (Cu) anode of 1.5406 Å X-ray radiation. A scanning electron microscope (SEM) LEO 1525 equipped with EDX was used to evaluate the general composition of KNN films and the overall morphology and macroscopic segregation shown in the Supporting Information. A field emission electron microscope Hitachi SU8220 with higher resolution was used to image the surface and cross-section of the films. Grain size measurements were performed on the high-resolution SEM images using the Gwyddion software. A THEMIS Z TEM of ThermoFisher was used to characterize the morphology and local composition of KNN films. The lamellas were taken approximately from the middle of the cross-section of KNN films. Raman spectroscopy was performed using a Renishaw inVia system equipped with a 20× objective. A 514 nm (green) laser was used as excitation source with 10% power. Spectra were acquired with a 2400 lines mm$^{-1}$ grating, an integration time of 20 s per spectrum, and 5 accumulations.

To assess the ferroelectric properties and leakage currents in KNN films, square platinum (Pt) contacts measuring 200 μm x 200 μm were fabricated using maskless lithography and ion beam etching. The ferroelectric properties were characterized using an aixACCT's TF Analyzer 2000 instrument. The piezoelectric and dielectric properties were measured by Double Beam Laser Interferometer (DBLI) with aixDBLI from aixACCT Systems GmbH. Piezoelectric properties were assessed by pattering top Platinum (Pt) electrodes with square (from 20 μm to 400 μm of lateral dimension) and circular (from 20 μm to 960 μm of lateral dimension) shapes with different sizes. The dielectric constant was estimated from capacitance measurements on several (three to four) capacitors, having different areas, for each measurement site. The values were obtained by linearly fitting the capacitance vs the capacitor area, taking the slope of the obtained line, and multiplying by a factor $d/\varepsilon_0$, where $d$ is the film thickness and $\varepsilon_0$ is the dielectric constant of vacuum. This procedure is useful to exclude parasitic or spurious contributions to the read capacitance, which may arise from the experimental setup. The true values of piezoelectric coefficients $e_{31,\mathrm{f}}$ and $d_{33,\mathrm{f}}$ were determined from DBLI measurements of the apparent $d_{33\mathrm{f,meas}}$ on capacitors with different areas. The coefficients were obtained from the well-known method of fitting the $d_{33\mathrm{f,meas}}$ vs the capacitor area.[67]

For the DFT calculations, we largely adopted the setup used in our previous work[11] to ensure methodological consistency. Simulations were performed within the Vienna Ab-initio Simulation Package (VASP)[20] using the PBEsol exchange-correlation functional.[68] Projector-augmented wave (PAW)[69] pseudopotentials were employed, with the following valence configurations: K (3s, 3p, 4s), Na (2s, 2p, 3s), Nb (4p, 5s, 4d), and O (2s, 2p). A plane-wave kinetic-energy cutoff of 800 eV was used for all systems (deemed necessary to converge total energy differences to within 1 meV unit cell$^{-1}$ in separate tests for the parent compound $KNbO_3$).

All simulations were carried out using a 2×3×2 supercell, constructed from the 10-atom monoclinic unit cell (space group $Pm$, tilt system $a^0b^+c^0$). This corresponds to a total of 120 atoms, of which 24 belong to the A-site sublattice; the Na concentration is then naturally defined as $x = N_{Na}/24$. The Brillouin zone was sampled with a 3×3×3 $k$-point mesh, corresponding to a 6×9×6 mesh for the primitive 10-atom cell. The chemical disorder on the A sublattice was modelled by means of the Special Quasirandom Structure (SQS) technique[70] as implemented in the software ICET.[71] Different SQS configurations were necessarily generated for each Na concentration. The cluster expansion, common to all systems, was restricted to pair interactions (cutoff 9 Å), yielding on the order of 40 effective cluster interactions.

Initial lattice parameters were generated according to Vegard's law, via linear interpolation (and suitable rescaling to the supercell size) between the equilibrium lattice constants of monoclinic $KNbO_3$ ($a_1$ = 5.698 Å, $a_2$ = 3.962 Å, $a_3$ = 5.682 Å) and $NaNbO_3$ ($a_1$ = 5.535 Å, $a_2$ = 3.920 Å, $a_3$ = 5.522 Å), obtained from independent calculations. All structures were subsequently fully relaxed. Convergence thresholds were set to $10^{-6}$ eV for electronic self-consistency and $10^{-4}$ eV for ionic relaxation, resulting in residual forces below 0.03 eV Å$^{-1}$. Piezoelectric tensors were computed using density-functional perturbation theory (DFPT), with the same electronic convergence criterion.

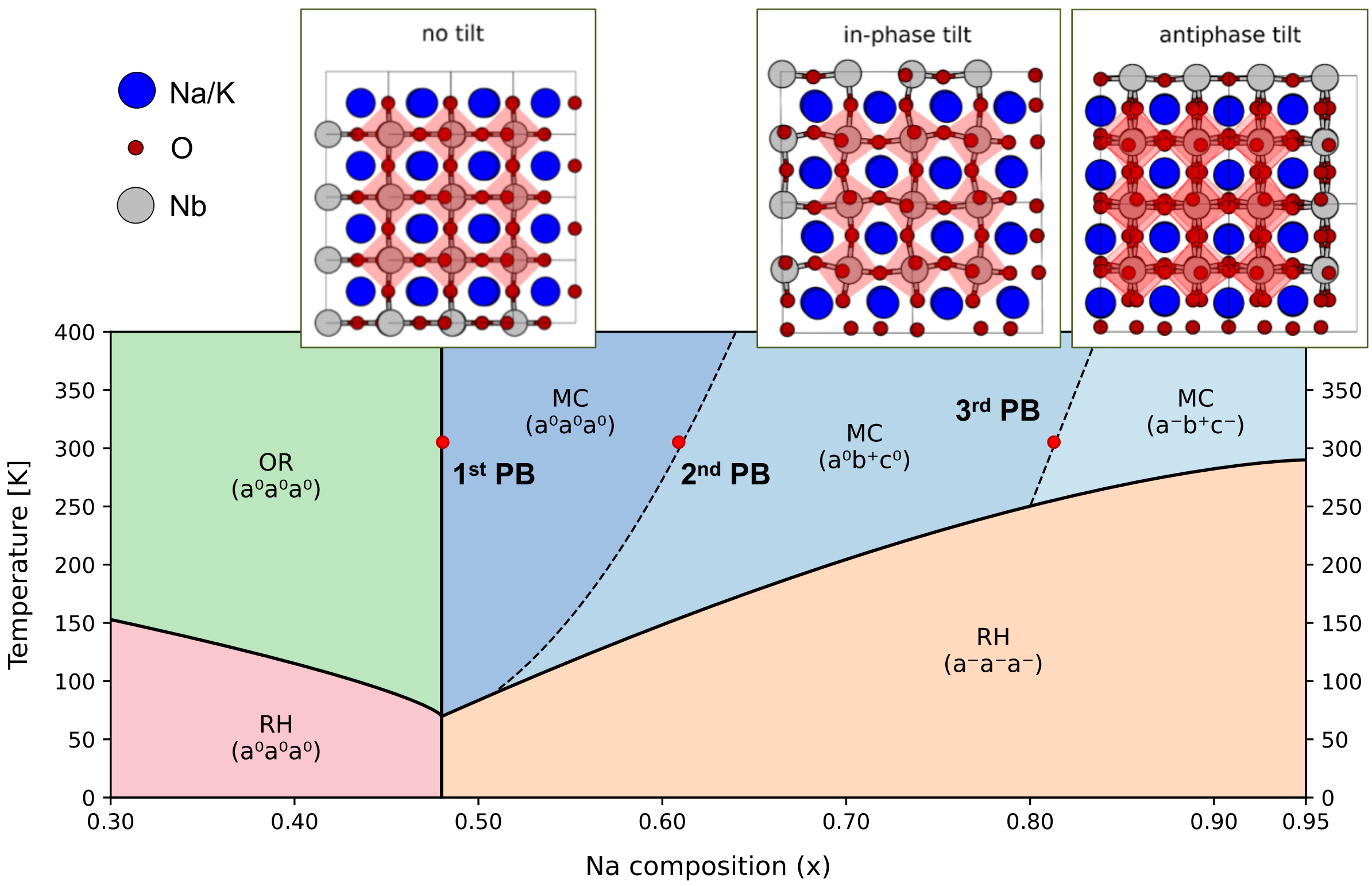

**Figure 1.** Phase diagram of KNN showing the stability regions of orthorhombic (OR), rhombohedral (RH), and monoclinic (MC) phases as a function of Na composition and temperature below 400 K (data replotted from Zhang et al.[20]). The labels in parentheses indicate the corresponding octahedral tilt systems in Glazer notation.[21] The atomistic models shown above the diagram illustrate, from left to right, untilted, in-phase tilted, and antiphase tilted configurations of the oxygen octahedra (drawn in light red), atop the phases where they first appear. Regions of phase transitions due to compositional differences have been marked as 1st PB, 2nd PB, and 3rd PB, where PB stands for phase boundary. These sketches serve only as schematic examples of the corresponding tilt systems; they are not intended as exact structural representations of the phases shown below, nor do they correspond in a one-to-one manner. Color code for the atomic species is: Na/K - blue, Nb - grey, O - red.

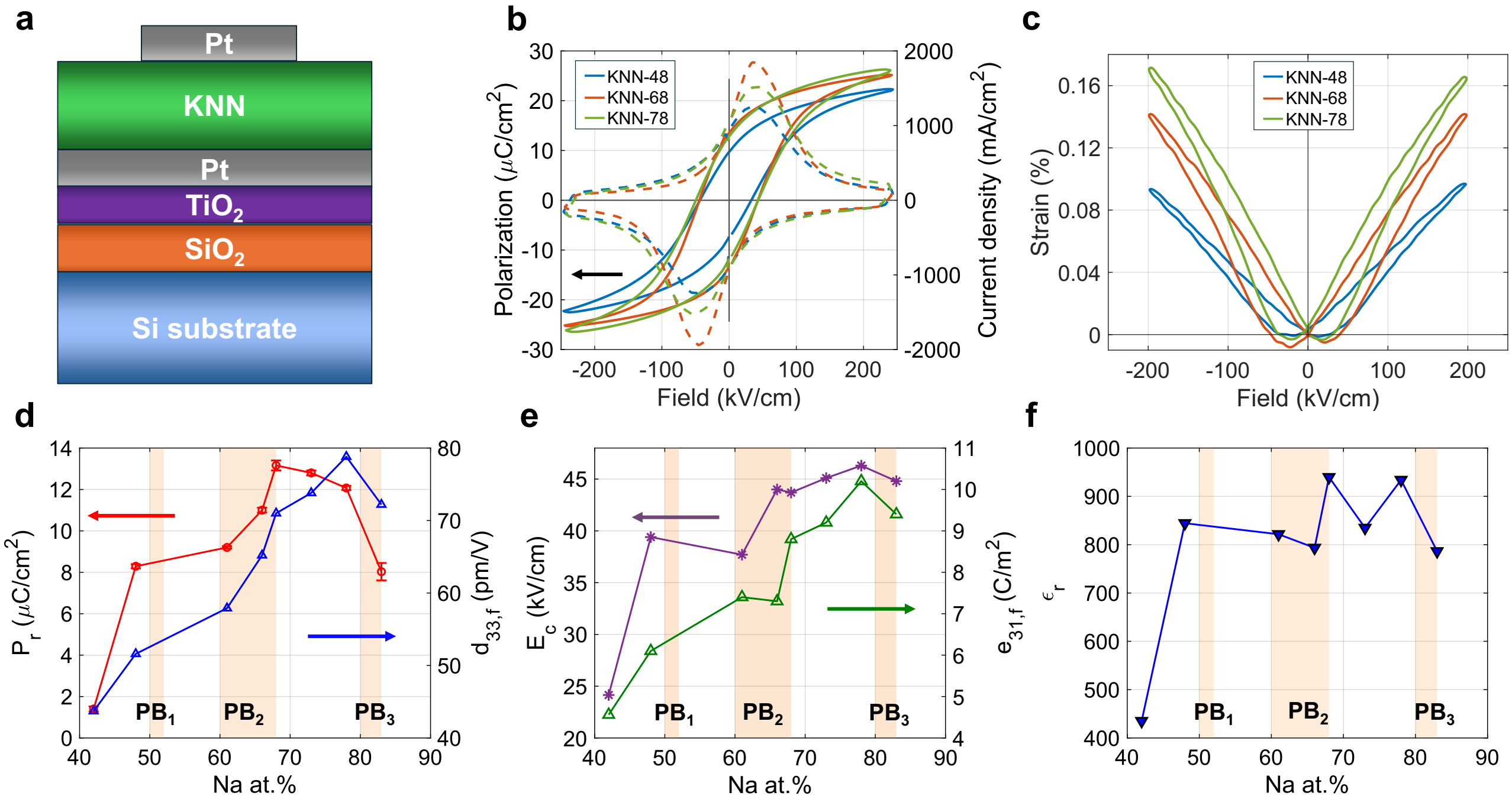

**Figure 2.** (a) Typical KNN device stack grown by co-sputtering, (b) polarization (current) loops, and (c) strain loops for three compositions near phase boundaries of KNN. (d) Remanent polarization, $d_{33,f}$ coefficients, (e) $e_{31,f}$ coefficients, coercive fields, and (f) trends for the dielectric constant for increasing Na compositions. The orange shaded regions in (d), (e), and (f) indicate the three phase boundary regions reported for bulk systems at room temperature ($PB_1$, $PB_2$, and $PB_3$). The values reported in (d) and (e) represent the average of the measurements obtained under positive and negative electric fields.

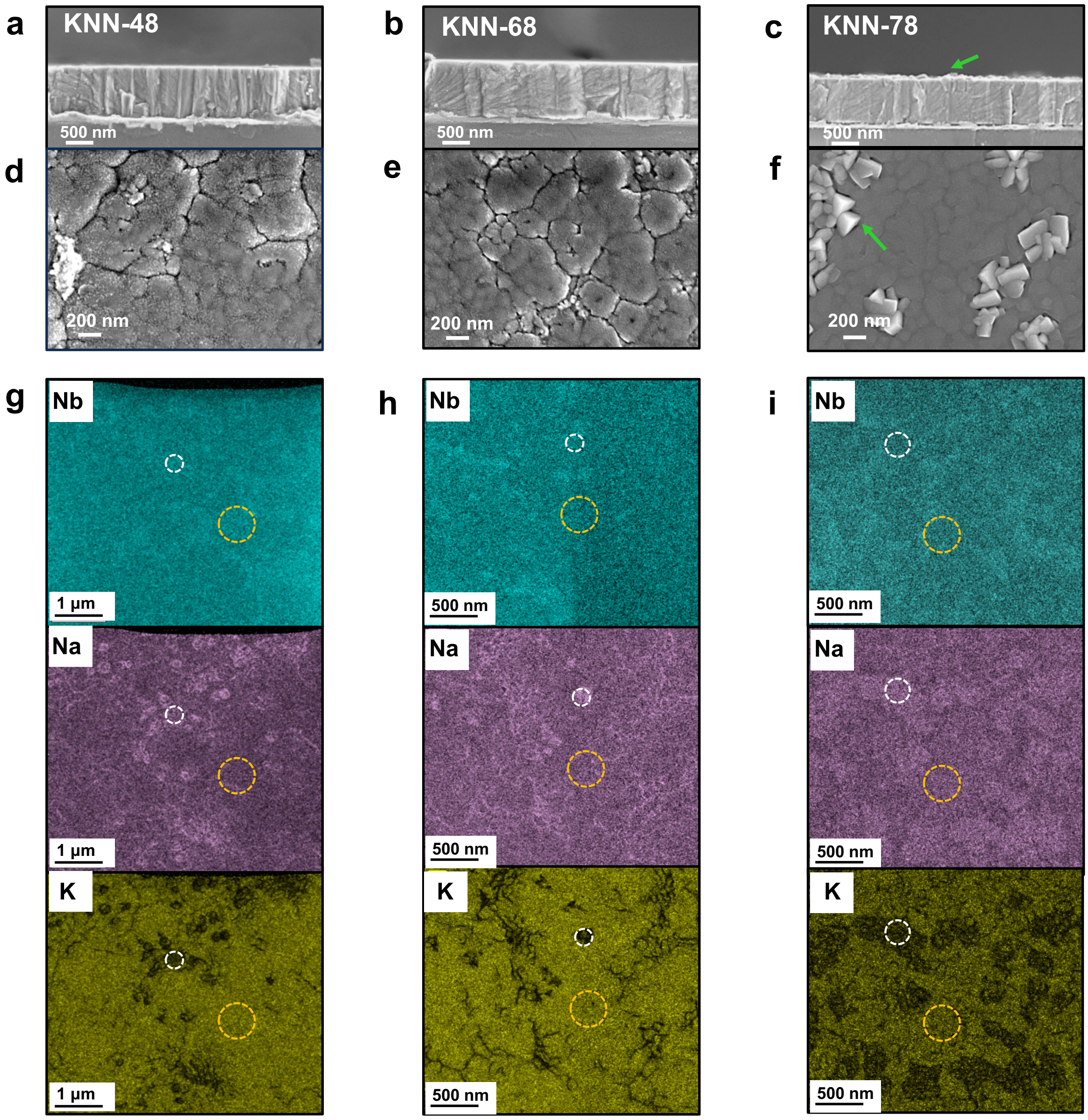

**Figure 3.** SEM cross section morphology and compositional STEM-EDX of (a, d, g) KNN-48, (b, e, h) KNN-68 and (c, f, i) KNN-78 films. The orange circles are used to denote even regions of uniform composition while the white circles underline film regions where Na is in higher proportion. The STEM-EDX images correspond to horizontal KNN lamellas (planar) taken from the middle of the films.

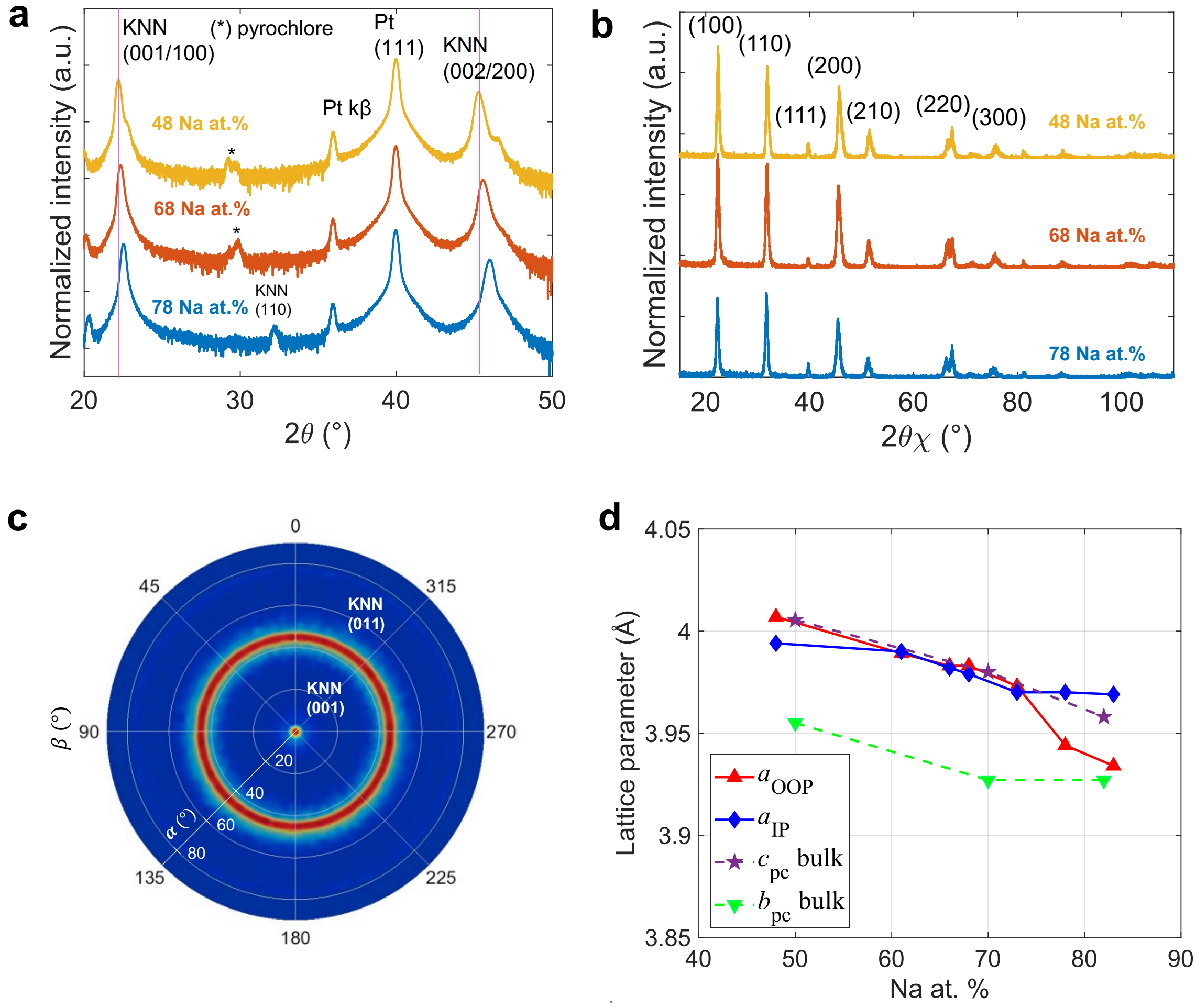

**Figure 4.** (a) Out of plane (θ-2θ), and (b) in-plane (2θχ) XRD diffractograms of KNN films with varying alkali compositions for the two measuring geometries. (c) Fused polar plots displaying KNN (001) in out of plane orientation, and a Debye-ring of KNN (011) at 45° degrees from the vertical, justifying a fiber texture quality. (d) Estimated out of plane ($a_{OOP}$) and in-plane ($a_{IP}$) lattice parameters as a function of Na at.% for KNN films and estimated lattice parameters $c_{pc}$ and $b_{pc}$ from KNN ceramics in the pseudo-cubic framework (from Ishizawa and Baker et al.[26,29,58]).

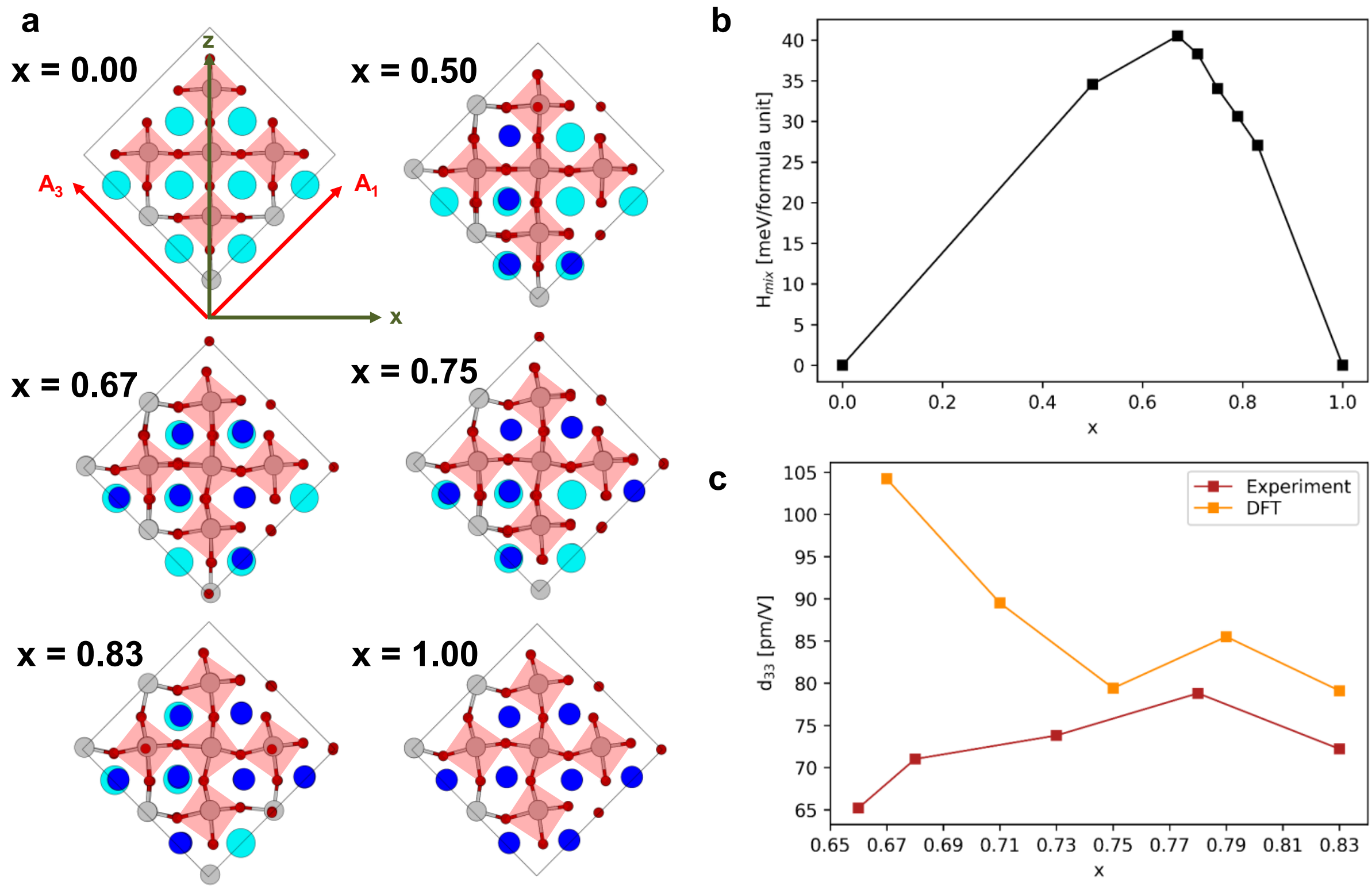

**Figure 5.** a) Views of the relaxed supercells (*xz* plane) for a representative subset of the KNN systems. Global *x* and *z* axes are drawn in green, supercell lattice vectors $A_1$ and $A_3$ in red (both *y* and $A_2$ point inwards, according to the right-hand rule). Color code for the atomic species is: Na - blue, K - turquoise, Nb - grey, O - red. Oxygen octahedra are drawn in light red. b) Mixing enthalpy $H_{\mathrm{mix}}$ of the $a^0b^+c^0$ phase, normalized per 5-atom perovskite formula unit, for various Na compositions. c) Comparison between the $d_{33}$ coefficients evaluated via DFT for the monoclinic $a^0b^+c^0$ phase of bulk KNN, as a function of the Na composition, and the experimental values measured on KNN films.

## References

[1] European Parliament and Council, Directive 2011/65/EU on the Restriction of the Use of Certain Hazardous Substances in Electrical and Electronic Equipment (RoHS), Off. J. Eur. Union, L174, 88, **2011**.
[2] A. Behera, in Advanced Materials: An Introduction to Modern Materials Science (Ed.: A. Behera), Springer International Publishing, Cham, **2022**, pp. 43–76.
[3] H. Hoshyarmanesh, M. Ghodsi, M. Kim, H. H. Cho, H.-H. Park, H. Hoshyarmanesh, M. Ghodsi, M. Kim, H. H. Cho, H.-H. Park, Sensors **2019**, 19, DOI 10.3390/s19122805.
[4] H. Kulkarni, K. Zohaib, A. Khusru, K. Shravan Aiyappa, Materials Today: Proceedings **2018**, 5, 21299.
[5] J. Liu, H. Tan, X. Zhou, W. Ma, C. Wang, N.-M.-A. Tran, W. Lu, F. Chen, J. Wang, H. Zhang, J. Appl. Phys. **2025**, 137, 020702.
[6] Y. Saito, H. Takao, T. Tani, T. Nonoyama, K. Takatori, T. Homma, T. Nagaya, M. Nakamura, Nature **2004**, 432, 84.
[7] J. Cai, Z. Cen, C. Zhou, N. Luo, Z. Dou, D. Yu, Z. Xu, ACS Appl. Energy Mater. **2025**, 8, 11393.
[8] K. Xu, J. Li, X. Lv, J. Wu, X. Zhang, D. Xiao, J. Zhu, Advanced Materials **2016**, 28, 8519.
[9] C. W. Ahn, S. Y. Lee, H. J. Lee, A. Ullah, J. S. Bae, E. D. Jeong, J. S. Choi, B. H. Park, I. W. Kim, J. Phys. D: Appl. Phys. **2009**, 42, 215304.
[10] C. Groppi, F. Maspero, M. Asa, G. Pavese, C. Rinaldi, E. Albisetti, M. Badillo-Avila, R. Bertacco, Journal of Applied Physics **2023**, 134, 204102.
[11] G. Pavese, F. Orlando, S. Picozzi, E. Albisetti, F. Maspero, M. Asa, F. Melzi, L. Castoldi, R. Bertacco, M. Badillo, Journal of Alloys and Compounds **2025**, 1022, 180039.
[12] K. Shibata, K. Watanabe, T. Kuroda, T. Osada, Appl. Phys. Lett. **2022**, 121, 092901.
[13] "KNN Lead-free Piezoelectric Thin Films | Products | SCIOCS," can be found under https://www.sumitomo-chem.co.jp/sciocs/english/products/KNN.html, **2026**.
[14] H. M. A. Piah, M. W. A. Rashid, U. A.-A. Azlan, M. A. M. Hatta, H. M. A. Piah, M. W. A. Rashid, U. A.-A. Azlan, M. A. M. Hatta, AIMSMATES **2023**, 10, 835.
[15] J. E. Garcia, F. Rubio-Marcos, J. Appl. Phys. **2020**, 127, 131102.
[16] M. Ahart, M. Somayazulu, R. E. Cohen, P. Ganesh, P. Dera, H. Mao, R. J. Hemley, Y. Ren, P. Liermann, Z. Wu, Nature **2008**, 451, 545.
[17] Z. Q. Zhuang, M. J. Haun, S.-J. Jang, L. E. Cross, IEEE Transactions on Ultrasonics, Ferroelectrics, and Frequency Control **1989**, 36, 413.
[18] B. Noheda, J. A. Gonzalo, L. E. Cross, R. Guo, S.-E. Park, D. E. Cox, G. Shirane, Phys. Rev. B **2000**, 61, 8687.
[19] B. Jaffe, R. S. Roth, S. Marzullo, J. Appl. Phys. **1954**, 25, 809.
[20] G. Kresse, J. Furthmüller, Phys. Rev. B **1996**, 54, 11169.
[21] A. M. Glazer, Acta Cryst B **1972**, 28, 3384.
[22] D. W. Baker, P. A. Thomas, N. Zhang, A. M. Glazer, Appl. Phys. Lett. **2009**, 95, DOI 10.1063/1.3212861.
[23] M.-H. Zhang, C. Shen, C. Zhao, M. Dai, F.-Z. Yao, B. Wu, J. Ma, H. Nan, D. Wang, Q. Yuan, L. L. da Silva, L. Fulanović, A. Schökel, P. Liu, H. Zhang, J.-F. Li, N. Zhang, K. Wang, J. Rödel, M. Hinterstein, Nat Commun **2022**, 13, 3434.
[24] M. Ahtee, A. W. Hewat, Acta Cryst A **1978**, 34, 309.
[25] N. Zhang, A. M. Glazer, D. Baker, P. A. Thomas, Acta Cryst B **2009**, 65, 291.
[26] N. Ishizawa, J. Wang, T. Sakakura, Y. Inagaki, K. Kakimoto, Journal of Solid State Chemistry **2010**, 183, 2731.

[27] I. Levin, V. Krayzman, G. Cibin, M. G. Tucker, M. Eremenko, K. Chapman, R. L. Paul, Sci Rep **2017**, 7, 15620.
[28] Y.-J. Dai, X.-W. Zhang, K.-P. Chen, Applied Physics Letters **2009**, 94, 042905.
[29] D. W. Baker, P. A. Thomas, N. Zhang, A. M. Glazer, Applied Physics Letters **2009**, 95, 091903.
[30] B.-P. Zhang, J.-F. Li, K. Wang, H. Zhang, Journal of the American Ceramic Society **2006**, 89, 1605.
[31] J. Rodel, W. Jo, K. Seifert, E.-M. Anton, T. Granzow, D. Damjanovic, Journal of the American Ceramic Society **2009**, 92, DOI 10.1111/j.1551-2916.2009.03061.x.
[32] L. Egerton, D. M. Dillon, Journal of the American Ceramic Society **1959**, 42, 438.
[33] H. E. Mgbemere, R.-P. Herber, G. A. Schneider, Journal of the European Ceramic Society **2009**, 29, 3273.
[34] G.-Z. Zang, X.-J. Yi, J. Du, Y.-F. Wang, Materials Letters **2010**, 64, 1394.
[35] Y. Noguchi, M. Miyayama, Journal of the Ceramic Society of Japan **2010**, 118, 711.
[36] D. G. Schlom, L.-Q. Chen, C.-B. Eom, K. M. Rabe, S. K. Streiffer, J.-M. Triscone, Annual Review of Materials Research **2007**, 37, 589.
[37] I. Kanno, T. Mino, S. Kuwajima, T. Suzuki, H. Kotera, K. Wasa, IEEE Transactions on Ultrasonics, Ferroelectrics, and Frequency Control **2007**, 54, 2562.
[38] J. Huang, J. Liu, Z. Li, K. Zhu, B. Wang, Q. Gu, B. Feng, J. Qiu, J Mater Sci: Mater Electron **2016**, 27, 899.
[39] L. Wang, W. Ren, W. Ma, M. Liu, P. Shi, X. Wu, AIP Advances **2015**, 5, 097120.
[40] L. Xu, Z. Kong, B. Zhu, X. Wang, K. Han, P. Chen, C. Li, W. Wu, F.-Z. Yao, K. Wang, Z. Huang, F. Chen, Nat Commun **2025**, 16, 5907.
[41] T. Matsuda, W. Sakamoto, B.-Y. Lee, T. Iijima, J. Kumagai, M. Moriya, T. Yogo, Jpn. J. Appl. Phys. **2012**, 51, 09LA03.
[42] N. D. Phu, X. L. Le, N. X. Duong, J Mater Sci: Mater Electron **2024**, 35, 1564.
[43] X. Zhang, J. Liu, K. Zhu, J. Wang, Z. Li, J. Qiu, J Mater Sci: Mater Electron **2017**, 28, 487.
[44] F. Fu, B. Shen, J. Zhai, Z. Xu, X. Yao, Ceramics International **2012**, 38, S287.
[45] N. D. Phu, X. L. Le, N. X. Duong, J Mater Sci: Mater Electron **2024**, 35, 1564.
[46] J. Wu, J. Wang, J. Appl. Phys. **2009**, 106, 066101.
[47] K.-N. Pham, N. H. Gaukås, M. Morozov, T. Tybell, P. E. Vullum, T. Grande, M.-A. Einarsrud, R Soc Open Sci. **2019**, 6, 180989.
[48] N. Kondo, W. Sakamoto, B.-Y. Lee, T. Iijima, J. Kumagai, M. Moriya, T. Yogo, Jpn. J. Appl. Phys. **2010**, 49, 09MA04.
[49] C. Groppi, L. Mondonico, F. Maspero, C. Rinaldi, M. Asa, R. Bertacco, Journal of Applied Physics **2021**, 129, 194102.
[50] A. S. Karapuzha, N. K. James, H. Khanbareh, S. van der Zwaag, W. A. Groen, Ferroelectrics **2016**, 504, 160.
[51] Y. Cao, G. Sheng, J. X. Zhang, S. Choudhury, Y. L. Li, C. A. Randall, L. Q. Chen, Appl. Phys. Lett. **2010**, 97, 252904.
[52] K. H. Yoon, H. R. Lee, Journal of Applied Physics **2001**, 89, 3915.
[53] P. Kumar, M. Pattanaik, Sonia, Ceramics International **2013**, 39, 65.
[54] Y.-J. Dai, X.-W. Zhang, K.-P. Chen, Appl. Phys. Lett. **2009**, 94, 042905.
[55] M. D. Nguyen, E. P. Houwman, M. Dekkers, G. Rijnders, ACS Appl. Mater. Interfaces **2017**, 9, 9849.
[56] K. Shibata, F. Oka, A. Nomoto, T. Mishima, I. Kanno, Jpn. J. Appl. Phys. **2008**, 47, 8909.
[57] D. Nečas, P. Klapetek, centr.eur.j.phys. **2012**, 10, 181.
[58] D. W. Baker, P. A. Thomas, N. Zhang, A. M. Glazer, Acta Cryst B **2009**, 65, 22.
[59] M. D. Nguyen, E. P. Houwman, G. Rijnders, Sci Rep **2017**, 7, 12915.

[60] R. D. Shannon, Acta Cryst A **1976**, 32, 751.
[61] G. C. A. M. Janssen, M. M. Abdalla, F. van Keulen, B. R. Pujada, B. van Venrooy, Thin Solid Films **2009**, 517, 1858.
[62] K. Suenaga, K. Shibata, K. Watanabe, A. Nomoto, F. Horikiri, T. Mishima, Jpn. J. Appl. Phys. **2010**, 49, 09MA05.
[63] H. Fu, R. E. Cohen, Nature **2000**, 403, 281.
[64] B. Noheda, D. E. Cox, G. Shirane, R. Guo, B. Jones, L. E. Cross, Phys. Rev. B **2000**, 63, 014103.
[65] P. C. Goh, K. Yao, Z. Chen, Journal of the American Ceramic Society **2009**, 92, 1322.
[66] H. Bruncková, Ľ. Medvecký, P. Hvizdoš, Materials Science and Engineering: B **2013**, 178, 254.
[67] S. Sivaramakrishnan, P. Mardilovich, T. Schmitz-Kempen, S. Tiedke, Journal of Applied Physics **2018**, 123, 014103.
[68] J. P. Perdew, A. Ruzsinszky, G. I. Csonka, O. A. Vydrov, G. E. Scuseria, L. A. Constantin, X. Zhou, K. Burke, Phys. Rev. Lett. **2008**, 100, 136406.
[69] P. E. Blöchl, Phys. Rev. B **1994**, 50, 17953.
[70] A. Zunger, S.-H. Wei, L. G. Ferreira, J. E. Bernard, Phys. Rev. Lett. **1990**, 65, 353.
[71] M. Ångqvist, W. A. Muñoz, J. M. Rahm, E. Fransson, C. Durniak, P. Rozyczko, T. H. Rod, P. Erhart, Advanced Theory and Simulations **2019**, 2, 1900015.

**Funding**

The authors report financial support from STMicroelectronics s.r.l. during the study. The co-author from IMRE received support from A*STAR, RIE2025, IAF-ICP Grant I2301E0027.

**Acknowledgements**

This work was supported by the Joint Research Centre STEAM between Politecnico di Milano and STMicroelectronics. The authors acknowledge the availability of experimental facilities at PoliFAB, and at IMRE, A*STAR, with acknowledge of support from A*STAR RIE2025, IAF-ICP Grant I2301E0027. We would like to express our gratitude to David Lim Boon Kiang for his help with high resolution SEM imaging.

**Data Availability Statement**

The data that support the findings of this study are available from the corresponding authors upon reasonable request.